\documentclass[pra,reprint,twocolumn,superscriptaddress,showpacs,floatfix]{revtex4-1}
\usepackage{amsmath}
\usepackage{mathrsfs}
\usepackage{txfonts}
\usepackage{amssymb}
\usepackage{graphicx}
\usepackage{hyperref}
\usepackage{ulem}
\usepackage{overpic}
\usepackage{psfrag}
\usepackage{tabularx}
\usepackage{array}
\usepackage{placeins}
\newcommand{\PreserveBackslash}[1]{\let\temp=\\#1\let\\=\temp}

\makeatletter

\begin{document}

\title{Spontaneous symmetry breaking with type-B Goldstone modes\\ in the SO($2s+1$) ferromagnetic model: an entanglement perspective}

\author{Qian-Qian Shi}
\affiliation{Centre for Modern Physics, Chongqing University, Chongqing 400044, The People's Republic of China}

\author{Huan-Qiang Zhou}
\affiliation{Centre for Modern Physics, Chongqing University, Chongqing 400044, The People's Republic of China}

\author{Murray T. Batchelor}
\affiliation{Mathematical Sciences Institute, The Australian National University, Canberra ACT 2601, Australia}
\affiliation{Centre for Modern Physics, Chongqing University, Chongqing 400044, The People's Republic of China}

\author{Ian P. McCulloch}
\affiliation{Department of Physics, National Tsing Hua University, Hsinchu 30013, Taiwan}
\affiliation{Frontier Center for Theory and Computation, National Tsing Hua University, Hsinchu 30013, Taiwan}

\begin{abstract}
Spontaneous symmetry breaking with type-B Goldstone modes is investigated in the SO($2s+1$) ferromagnetic model. 
A set of orthonormal basis states in the ground state subspace are constructed, which admit an exact Schmidt decomposition, exposing self-similarities in real space of an abstract fractal underlying the ground state subspace.   
Focusing on the SO(5) and the SO(6) ferromagnetic spin chains as illustrative examples, finite system-size scaling analysis of the  entanglement entropy for this set of orthonormal basis states confirms that the entanglement entropy scales logarithmically with block size in the thermodynamic limit. The prefactor in front of the logarithm is half the number of type-B Goldstone modes $N_B$, which is identified as the fractal dimension $d_f$ for these orthonormal basis states. For the SO($2s+1$) ferromagnetic model $N_B = d_f =s$ for integer $s$ and $N_B = d_f =s+1/2$ for half-odd-integer $s$. 
\end{abstract}
\maketitle

\section{Introduction}
The ${\rm SU}(2)$ spin-1 bilinear-biquadratic model continues to attract attention as a key model in the realm of quantum many-body physics (see, e.g., Refs~\cite{Chubukov, Fath, Fath2, Kawashima, Ivanov,  Buchta, Rizzi, Lauchli, Porras, Romero, ronny,lundgren, Rakov, Sierra, daibb}). 
The physical richness of this model relates to the fact that it includes several remarkable quantum many-body systems as special cases. 
These specific models occupy a prominent position in both condensed matter and mathematical physics, being relevant to the quantum Yang-Baxter equation~\cite{baxterbook,sutherlandb,mccoy}, the Temperley-Lieb algebra~\cite{tla,baxterbook,martin}, the non-linear $\sigma$ model~\cite{nonlinear}, frustration-free models~\cite{tasaki,katsura} and conformal field theory~\cite{cft}. 
Examples are the SU(2) spin-1 antiferromagnetic Heisenberg model~\cite{Haldane} and the
Affleck-Kennedy-Lieb-Tasaki model~\cite{AKLT}, which are adiabatically connected to each other, thus solving some of the initial mystery around the celebrated Haldane gap~\cite{Haldane,Oshikawa}. Other examples include the SU(3) Uimin-Lai-Sutherland model~\cite{Sutherland}, the SU(2) spin-1 Takhtajan-Babujian model~\cite{TB}, and the SU(3) spin-1 biquadratic model~\cite{Parkinson, Barber, Klumper}. All reveal many surprising connections between diverse areas in physics and mathematics.

Recently, one of the unexpected connections concerns a variety of exactly solvable models involving ferromagnetic interactions, the symmetry group ${\rm SU}(N)$ and spontaneous symmetry breaking (SSB) with type-B Goldstone modes (GMs)~\cite{FMGM,LLspin1,golden,SU4,finitesize}. 
The latter is part of the sustained effort towards a proper classification of GMs~\cite{goldstone,Hnielsen, schafer, miransky, nambu, nicolis,  brauner-watanabe, watanabe, NG}, culminating in the introduction of type-A and type-B GMs~\cite{watanabe}.  
It is found that there are a set of orthonormal basis states in the ground-state subspace for a quantum many-body  system undergoing SSB with type-B GMs. They are subject to exact Schmidt decomposition, revealing self-similarities in real space, leading in turn to an abstract fractal underlying the ground state subspace, as demonstrated in Ref.~\cite{cantor} for the SU(2) spin-$s$ ferromagnetic Heisenberg model and the SU($2s+1$) ferromagnetic model.
As a result, the number of type-B GMs is identified  with the fractal dimension - a mathematical notion first exploited in Ref.~\cite{doyon} and further developed in Ref.~\cite{cantor}, as far as the orthonormal basis states are concerned. 
As a consequence of the presence of exact Schmidt decompositions,  
the highly degenerate ground states are expressible in terms of an {\it exact} matrix product state (MPS) representation, which in turn makes it possible to reveal an {\it emergent} cyclic permutation symmetry group in characterization of this type of scale invariant states.

The appearance of type-B GMs has also recently been explored in the study of SSB, entanglement and logarithmic spirals in a quantum spin-1 many-body system with competing dimer and trimer interactions~\cite{DTmodel}. 
For the one-dimensional flat-band ferromagnetic Tasaki model entanglement entropy scaling reveals SSB with one type-B GM~\cite{Tasakimodel}.

In the present work we explore ferromagnetic interactions with a different symmetry group. 
Given the isomorphism between the Lie algebras of  SU(2) and SO(3), a natural candidate for investigation is the known extension of the SU(2) spin-1 bilinear-biquadratic model to the SO($2s+1$) bilinear-biquadratic model~\cite{son,alet,orus,okunishi,Hakobyan,chulliparambil}:
\begin{equation}
		\mathscr{H}=\cos\theta\sum_i\sum_{a<b} L_i^{ab}L_{i+1}^{ab}+\sin\theta\sum_i\left(\sum_{a<b}L_i^{ab}L_{i+1}^{ab}\right)^2 \,.
	\label{ham}
\end{equation}
Here $L_{ab}(1\leq a<b\leq 2s+1)$ denotes the $s(2s+1)$ generators of the symmetry group SO($2s+1$), which transform in the vectorial representation. Note that the dimension of the local Hilbert space is $2s+1$, if the model is realized in terms of the spin-$s$ operators~\cite{son,alet}. Thus the Hamiltonian (\ref{ham}) represents a family of spin-$s$ quantum many-body systems, with the coupling constants parametrized by $\theta\in [-\pi,\pi]$.

A ground state phase diagram~\cite{son} has been sketched from a few characteristic points that are exactly solvable, including the SU($2s+1$) Uimin-Lai-Sutherland model~\cite{Sutherland} 
located at $\theta=\tan ^{-1}[1/(2s-1)]$, the frustration-free Affleck-Kennedy-Lieb-Tasaki (AKLT) model~\cite{son,AKLT,Kolezhuk-Scalapino} located at $\theta=\tan^{-1}[1/(2s+1)]$, 
the staggered SU($2s+1$) antiferromagnetic and ferromagnetic biquadratic models located at 
$\theta=\pm \pi/2$~\cite{Barber,Parkinson, Klumper, Klumper2}, 
and the Reshetikhin model~\cite{Reshetikhin} located at $\theta=\tan^{-1}[(2s-3)/(2s-1)^2]$. 
Indeed, the frustration-free AKLT model admits a ground state in a MPS representation, 
which behaves differently for half-odd-integer $s$ and integer $s$, namely a unique and translation-invariant ground state for integer $s$ and a two-fold degenerate ground state arising from spontaneously broken one-site translational symmetry for half-odd-integer $s$. 
Further, the staggered SU($2s+1$) antiferromagnetic biquadratic model may be mapped to the $(2s+1)^2$-state quantum Potts model~\cite{Batchelor}, 
and the Reshetikhin model is critical, described by the Wess-Zumino-Witten conformal field theories. 

In the present study, we investigate SSB with type-B GMs in the SO($2s+1$) ferromagnetic model, corresponding to Hamiltonian (\ref{ham}) at $\theta=\pi$. 
In particular, we focus on the SO(5) and SO(6)  ferromagnetic models as two representatives cases, in order to perform a systematic finite system-size scaling analysis for the  entanglement entropy. 
As a common feature in both cases, a set of orthonormal basis states are constructed and they admit an exact Schmidt decomposition, exposing an abstract fractal underlying the ground state subspace. 
The entanglement entropy is seen to scale logarithmically with the block size $n$ as the thermodynamic limit is approached. The prefactor in front of the logarithm is half the number of type-B GMs for the orthonormal basis states, which in turn is identified to be the fractal dimension. 
These numerical results provide further confirmation of the recent theoretical prediction~\cite{FMGM}.

The article is  organized as follows. In Section~\ref{ssb}, a SSB pattern for the SO(2s+1) ferromagnetic model is discussed, with a focus on  the SO(5) and the SO(6) ferromagnetic models. In Section~\ref{statesvd} a sequence of degenerate ground states arising from the highest weight state is presented for the SO(2s+1) ferromagnetic model, 
which form a set of orthonormal basis states in the ground state subspace. 
In Section~\ref{svdsm} an exact Schmidt decomposition is performed for the orthonormal basis states. Details of the expected entanglement entropy scaling are given in Section~\ref{NBentropy}. A universal finite system-size scaling analysis is then performed in Section~\ref{results} for the SO(5) and SO(6) cases. As a result, the identification of the fractal dimension with the number of type-B GMs is made. Section~\ref{summary} is devoted to a summary and concluding remarks.
  
\section{Spontaneous symmetry breaking in the SO($2s+1$) ferromagnetic model}\label{ssb}

We begin by devoting some attention to 
the symmetry group SO($2s+1$), which has a total of $s(2s+1)$ generators. 
When $s$ is an integer, the rank $r$ is equal to $s$. 
It is described by $s$ (commuting) Cartan generators $H_\alpha$ ($\alpha=1$, \ldots, $s$).
The rank $r$ is equal to $s+1/2$ if $s$ is half-odd-integer.
The description is then by $s+1/2$ (commuting) Cartan generators $H_\alpha$ ($\alpha=1$, \ldots, $s+1/2$).
For each $H_\alpha$, there exists a conjugate pair of a raising operator $E_\alpha$ and a lowering operator $F_\alpha$ such that $[H_\alpha, E_\beta]=\beta_\alpha E_\beta$, $[E_\alpha,F_\alpha]=(E_\alpha,F_\alpha)H_\alpha$, $[F_\alpha,H_\beta]=\beta_\alpha F_\beta$,  $[E_\alpha,E_\beta]=g_{\alpha,\;\beta}E_\gamma\delta_{\gamma,\;\alpha+\beta}$, and $[F_\alpha,F_\beta]=g_{-\alpha,\;-\beta}F_\gamma\delta_{\gamma,\;\alpha+\beta}$, with $\beta$ the root matrix, $g_{\alpha,\;\beta}$ depending on the specific form of the Cartan generators, and $(E_\alpha,F_\alpha)$ the Killing form of $E_\alpha$ and $F_\alpha$, which constitute $r$ ${\rm SU(2)}$ subgroups. 
In addition,  more lowering and raising generators, denoted by $E_\beta$ and  $F_\beta$, with $\beta=s+1, \ldots, s^2$ when $s$ is an integer and with $\beta=s+3/2,\ldots, s^2-1/4$ when $s$ is a half-odd-integer.  Here, we stress that it is convenient to choose the Cartan generators $H_\alpha$ in such a way that the set of the lowering operators $F_\alpha$ commute with each other, as far as the construction of a complete set of linearly independent degenerate ground states is concerned. This is due to the fact that, generically, degenerate ground states generated from the action of lowering operators on the (unique) highest weight state are amenable to a systematic finite system-size scaling analysis of the entanglement entropy for quantum many-body systems undergoing SSB with type-B GMs, if the symmetry group is semisimple.

To proceed, we focus on the SO(5) and SO(6) ferromagnetic models, in order to present the details of the construction in an explicit way. However, the extension to the SO($2s+1$) ferromagnetic model is straightforward. 

Specifically, for the symmetry group ${\rm SO}(5)$ with $s=2$, the total number of the generators is ten, with rank $r=2$.  
We may choose the two Cartan generators $H_1$ and $H_2$, with the two raising operators $E_1$ and $E_2$, and the two lowering operators $F_1$ and $F_2$ being determined accordingly. 
In addition,  there are four more generators, denoted by $E_3$, $E_4$,  $F_3$ and $F_4$.
The specific forms of the ten generators are listed in Sec.~A of the Appendix. 
If the highest weight state  $|\rm{hws}\rangle$ is chosen to be $|{\rm hws}\rangle=\vert\otimes_{k=1}^{L}  \{v_1\}_{\;k}\rangle$, with $|v_1\rangle=\sqrt{2}/2(0,1,0,\mathrm{i},0)^T$,
then the expectation values of the local components for the Cartan generators $H_{1,j}$ and $H_{2,j}$ are given by
$\langle v_1  | \, H_{1,j} \, | v_1\rangle=1/2$, $\langle v_1  | \, H_{2,j} \, |v_1\rangle=1/2$.
Here $v^T$ represents a vector transpose.  
For $E_1$, $E_2$, $F_1$ and $F_2$, one may choose $F_{\alpha,j}$ and $E_{\alpha,j}$ as the interpolating fields~\cite{nielsen, inter}.
Given that $\langle[E_\alpha,F_{\alpha,j}]\rangle\propto \langle H_{\alpha,j}\rangle$, $\langle[E_{\alpha,j},F_\alpha]\rangle\propto \langle H_{\alpha,j}\rangle$ ($\alpha=1, 2$), $\langle[E_3,F_{3,j}]\rangle\propto \langle H_{1,j}+H_{2,j}\rangle$, $\langle[E_{3,j},F_3]\rangle\propto \langle H_{1,j}+H_{2,j}\rangle$, 
$\langle[E_4,F_{4,j}]\rangle\propto \langle H_{1,j}-H_{2,j}\rangle$, and $\langle[E_{4,j},F_4]\rangle\propto \langle H_{1,j}-H_{2,j}\rangle$, we are led to conclude that the generators $E_\alpha$ and $F_\alpha$ ($\alpha=$1, 2, 3) are spontaneously broken, with $\langle H_{\alpha,j}\rangle$ ($\alpha=1$, and $2$) being the local order parameters. The SSB pattern is thus from SO(5) to  ${\rm U(1)}\times{\rm U(1)}$ via ${\rm U(1)}\times{\rm SU(2)}$ successively. 
That is, there are six broken generators. In contrast, there are only two (linearly) independent local order parameters, meaning that there are two type-B GMs: $N_B=2$.

For the symmetry group ${\rm SO}(6)$ with $s=5/2$, there are fifteen generators in total, with rank $r=3$. We remark that 
although  the Lie algebra ${\rm so}(6)$ is isomorphic to the Lie algebra ${\rm su}(4)$, the ${\rm SO}(6)$ ferromagnetic model is different from the  ${\rm SU}(4)$ ferromagnetic model, because two distinct irreducible representations are involved if one regards a representation space as a local Hilbert space at each lattice site.
It may be described by three Cartan generators $H_1$, $H_2$ and $H_3$,  with the raising operators $E_1$, $E_2$, $E_3$ and the lowering operators $F_1$, $F_2$, $F_3$.
In addition, there are six more generators, denoted by $E_4$, $F_4$,  $E_5$, $F_5$,  $E_6$ and $F_6$.
The specific form of these fifteen generators are also listed in Sec.~A of the Appendix. 
If the highest weight state  $|\rm{hws}\rangle$ is chosen to be $|{\rm hws}\rangle=\vert\otimes_{k=1}^{L}  \{v_1\}_{\;k}\rangle$, with $|v_1\rangle=\sqrt{2}/2(0,0,1,\mathrm{i},0,0)^T$, then the expectation values of the local components for the Cartan generators $H_{1,j}$, $H_{2,j}$ and $H_{3,j}$ are given by
$\langle v_1  | \, H_{1,j} \, | v_1\rangle=1/2$, $\langle v_1  | \, H_{2,j} \, |v_1\rangle=1/2$,
$\langle v_1  | \, H_{3,j} \, |v_1\rangle=1/2$. 
For $E_1$, $E_2$, $E_3$, $F_1$, $F_2$ and $F_3$, one may choose $F_{\alpha,j}$ and $E_{\alpha,j}$ as the interpolating fields~\cite{nielsen, inter}.
Given that $\langle[E_\alpha,F_{\alpha,j}]\rangle\propto \langle H_{\alpha,j}\rangle$, $\langle[E_{\alpha,j},F_\alpha]\rangle\propto \langle H_{\alpha,j}\rangle$ ($\alpha=1$, 2, 3), $\langle[E_4,F_{4,j}]\rangle\propto \langle H_{1,j}-H_{2,j}\rangle$, $\langle[E_{4,j},F_4]\rangle\propto \langle H_{1,j}-H_{2,j}\rangle$, 
$\langle[E_5,F_{5,j}]\rangle\propto \langle H_{3,j}\rangle$, $\langle[E_{5,j},F_5]\rangle\propto \langle H_{3,j}\rangle$, $\langle[E_6,F_{6,j}]\rangle\propto \langle H_{2,j}-H_{1,j}-H_{3,j}\rangle$, and $\langle[E_{5,j},F_5]\rangle\propto \langle H_{2,j}-H_{1,j}-H_{3,j}\rangle$ we are led to conclude that
the generators $E_\alpha$ and $F_\alpha$ ($\alpha=$1, 2, 3, 5, 6) are spontaneously broken, with $\langle H_{\alpha,j}\rangle$ ($\alpha=1$, and $2$) being the local order parameters. The SSB pattern is thus from SO(5)  to  ${\rm U(1)}\times{\rm U(1)}\times{\rm U(1)}$ via ${\rm U(1)}\times{\rm U(1)}\times{\rm SU(2)}$ successively. 
That is, there are ten broken generators. In contrast, there are only three (linearly) independent local order parameters, meaning that there are three type-B GMs: $N_B=3$.
 
Taking into account the Mermin-Wagner-Coleman theorem~\cite{mwc}, no type-A GM survives in one spatial dimension, thus the number of type-A GMs $N_A$ must be $N_A=0$.
As a consequence, there exists an apparent violation to the counting rule~\cite{watanabe} for the SSB pattern from SO(5) to $\rm{U(1)}\times\rm{U(1)}$ and from SO(5)  to  ${\rm U(1)}\times{\rm U(1)}\times{\rm U(1)}$ .

To retain consistency with the counting rule~\cite{watanabe}, we resort to the redundancy principle  introduced in Ref.~\cite{golden}.
For our purpose, the redundancy principle may be stated as follows. 
The number of GMs should not exceed the rank $r$ of the symmetry group $G$, as long as $G$ is semisimple.
For the symmetry group SO(5), the rank $r=2$. Hence the number of type-B GMs must be $N_B = 2$. For the symmetry group SO(6), the rank $r=3$. So the number of type-B GMs must be $N_B = 3$.
Two broken generators are thus redundant for the SSB pattern from SO(5) to ${\rm U(1)}\times{\rm U(1)}$ and four broken generators are redundant for the SSB pattern from SO(5)  to  ${\rm U(1)}\times{\rm U(1)}\times{\rm U(1)}$.

For the task at hand, we need to outline a general procedure to generate all of the degenerate ground states. Such a procedure consists of two steps as follows. The first step is to act with all the $r$ commuting lowering operators $F_\alpha$ ($\alpha=1$, \ldots, $r$) on the highest weight state $|\rm{hws}\rangle$. The second step is to act with a permutation symmetry operation $\Sigma$ on the degenerate ground states, already generated in the first step, to generate other degenerate ground states. Here $\Sigma$ is a (global) permutation symmetry operation  $\Sigma= \otimes_{j=1}^L \Sigma_j$, where $\Sigma_j$ acts on the $2s+1$ orthonormal basis states in the local Hilbert space at the $j$-th lattice site. Indeed, we have to deal with $(2s+1)!$ permutation operations that form a non-Abelian symmetric group $S_{2s+1}$. Note that $S_{2s+1}$ contains a subgroup $Z_2$ generated by a time-reversal operation.  In practice, it is not necessary to consider all possible choices of the $r$ commuting lowering operators in the first step, since {\it only} a few among all possible choices of the $r$ commuting lowering operators, together with $\Sigma$, are sufficient to generate all the linearly independent degenerate ground states. In particular, for the SO(5) and SO(6) ferromagnetic models, the ground state degeneracies are found to be given by 
\begin{align}
\mathrm{SO(5)}:& \quad (L+1)(L+2)(2L+3)/6, \\
\mathrm{SO(6)}:&  \quad (L+1)(L+2)^2(L+3)/12.
\end{align}

Actually for the ${\rm SO}(5)$ spin-$2$ case, there are four total choices for the two commuting lowering operators: $F_1$ and $F_2$, $F_1$ and $F_3$, $F_2$ and $F_3$, and $F_1$ and $F_4$.  We may choose two among the four choices: $F_1$ and $F_2$, and $F_1$ and $F_3$. 
For the ${\rm SO}(6)$ case, there are seven total choices of the three commuting lowering operators: $F_1$, $F_2$ and $F_3$, $F_1$, $F_2$ and $F_4$, $F_1$, $F_2$ and $F_6$, $F_1$, $F_3$ and $F_4$, $F_1$, $F_4$ and $F_5$,  $F_1$, $F_5$ and $F_6$, and $F_2$, $F_3$ and $F_4$.  
We may choose one among the seven choices: $F_1$, $F_2$ and $F_3$.
At this point, we emphasize that $\Sigma$ is (local) unitary so that its action does not change the Schmidt coefficients for a specific degenerate ground state, which in turn implies that the entanglement entropy remains to be identical for two degenerate ground states connected via $\Sigma$. Hence one {\it only} needs to focus on degenerate ground states generated from the highest weight state, as far as the Schmidt decomposition and the entanglement entropy are concerned.

The SO($2s+1$) ferromagnetic model may be realized in terms of spin-$s$ operators for $2s+1$ odd (integer $s$), but not for $2s+1$ even (half-odd-integer $s$). 
The ${\rm SO}(5)$ ferromagnetic model in terms of spin-$2$ operators is given in Sec.~B of the Appendix. %

\section{A set of orthonormal basis states as degenerate ground states}\label{statesvd}

For the SO($2s+1$) ferromagnetic model, a sequence of degenerate ground states $|L,M_1,\ldots,M_r\rangle$ are generated from the repeated action of the lowering operators $F_\alpha$ on the highest weight state $|\rm{hws}\rangle$ for $M_\alpha$ times with $\alpha=1,\ldots,r$:
\begin{equation}
|L,M_1,\ldots,M_r\rangle = 
\frac{1}{Z(L,M_1,\ldots,M_r)}\prod_{\alpha=1}^r F_\alpha^{\,\,M_\alpha}|\rm{hws}\rangle \,.
\label{lmnr}
\end{equation}
The term $Z(L,M_1,\ldots,M_r)$ is introduced to ensure that $|L,M_1,\ldots,M_r\rangle$ is normalized.
They form a set of orthonormal basis states in the ground state subspace, but not complete. This stems from the observation already made in the last Section that $|L,M_1,\ldots,M_r\rangle$ alone do not span the ground state subspace.

We again restrict ourselves to the ${\rm SO}(5)$ and ${\rm SO}(6)$ cases.

For the ${\rm SO}(5)$ ferromagnetic model, there are four pairs of two commuting lowering operators:  $F_1$ and $F_2$, $F_1$ and $F_3$, $F_2$ and $F_3$, and $F_1$ and $F_4$. Here we choose $F_1$ and $F_2$, and $F_1$ and $F_3$ to generate the degenerate ground states $|L,M_1,M_2\rangle$ ($M_1=0$, \ldots, $L$, $M_2=0$, \ldots, $L$), and $|L,M_1,M_3\rangle$ ($M_1=0$, \ldots, $L$, $M_3=0$, \ldots, $L$), respectively.  
For convenience, we denote $| \, L,M_1,M_3\rangle$ as $| \, L,M_1^*,M_2^*\rangle$, with $M_1^*=M_1$ and $M_2^*=M_3$. 
The fillings are defined as $f_1=M_1/L$ and $f_2=M_2/L$ or $f_1^*=M_1^*/L$ and $f_2^*=M_2^*/L$.

Note that the three local orthonormal states  $|v_2\rangle=\sqrt{2}/2(\mathrm{i},0,0,0,1)^T$, $|v_3\rangle=\sqrt{2}/2(\mathrm{i},0,0,0,-1)^T$ 
and $|v_4\rangle=\sqrt{2}/2(0,1,0,-\mathrm{i},0)^T$ are generated by acting with the lowering operators $F_{1,j}$ and $F_{2,j}$ on $|v_1\rangle$. That is, we have 
$F_{1,j}|v_1\rangle=|v_2\rangle$, $F_{2,j}|v_1\rangle=|v_3\rangle$, and $F_{1,j}F_{2,j}|v_1\rangle=|v_4\rangle$. Meanwhile they are the eigenvectors of the local components of the Cartan generators: $H_{1,j}|v_2\rangle=-1/2|v_2\rangle$, $H_{2,j}|v_2\rangle=1/2|v_2\rangle$, $H_{1,j}|v_3\rangle=1/2|v_3\rangle$,  $H_{2,j}|v_3\rangle=-1/2|v_3\rangle$, $H_{1,j}|v_4\rangle=-1/2|v_4\rangle$ and  $H_{2,j}|v_4\rangle=-1/2|v_4\rangle$.
In this way a sequence of the degenerate ground states $|L, M_1,M_2\rangle$ are generated from the repeated action of the lowering operators $F_1$ and $F_2$ on the highest weight state  $|{\rm hws}\rangle=|\otimes_{k=1}^{L} \{v_1\}_k\rangle$. Thus $|L,M_1,M_2\rangle=1/Z(L,M_1,M_2)F_1^{M_1}F_2^{M_2}|{\rm hws}\rangle$ may be rewritten as follows
\begin{widetext}
	\begin{equation}
		|L,M_1,M_2\rangle=\frac{1}{Z(L,M_1,M_2)}\sum_{P}{\sum}'_{N_1,N_2, N_3,N_4}
		|\underbrace{v_1v_1...}_{N_1}|\underbrace{v_2v_2...}_{N_2}|\underbrace{v_3v_3...}_{N_3}
		|\underbrace{v_4v_4...}_{N_4}\rangle,\label{lm1m2so5}
	\end{equation}
where the sum $\sum_{P}$ is taken over all permutations $P$ for a given partition $\{N_1,N_2, N_3, N_4\}$, and the sum ${\sum}'_{N_1,N_2, N_3,N_4}$ is taken over all possible values of $N_1,N_2, N_3$, and $N_4$, subject to the constraints $N_2+N_4=M_1$, $N_3+N_4=M_2$ and $N_1+N_2+N_3+N_4=L$.
Here $N_1,N_2,N_3$ and $N_4$ denote the numbers of the lattice sites in the configurations $|v_1\rangle$, $|v_2\rangle$, $|v_3\rangle$  and $|v_4\rangle$, respectively, and $Z(L,M_1,M_2)$ is introduced to ensure that $|L,M_1,M_2\rangle$ is normalized. This factor takes the form
\begin{equation}
	Z(L,M_1,M_2)=M_1!M_2!\sqrt{{\sum}'_{N_4}C_{L}^{N_4}C_{L-N_4}^{M_1-N_4}C_{L-M_1}^{M_2-N_4}}.
\end{equation}

Similarly the three local orthonormal states $v_2$, $v_4$ and $|v_5\rangle=(0,0,1,0,0)^T$  are generated by acting with the lowering operators $F_{1,j}$ and $F_{3,j}$ on $|v_1\rangle$. That is,  $F_{1,j}|v_1\rangle=|v_2\rangle$, $F_{3,j}|v_1\rangle=\sqrt{2}/2|v_5\rangle$ and  $(F_{3,j})^2|v_1\rangle=\sqrt{2}/2|v_4\rangle$.  Meanwhile, they are the eigenvectors of the local components of the Cartan generators: $H_{1,j}|v_5\rangle=0|v_5\rangle$ and  $H_{2,j}|v_5\rangle=0|v_5\rangle$, in addition to those for  $|v_2\rangle$ and $|v_4\rangle$, already discussed above.
A sequence of the degenerate ground states $|L,M_1^*,M_2^*\rangle$ are generated from the repeated action of the lowering operators $F_1$ and $F_3$ on the highest weight state  $|{\rm hws}\rangle=|\otimes_{k=1}^{L} \{v_1\}_k\rangle$.  
Now $|L,M_1^*,M_2^*\rangle=1/Z(L,M_1^*,M_2^*)F_1^{M_1}F_3^{M_2^*}|{\rm hws}\rangle$ may be rewritten as 
	\begin{equation}
		|L,M_1^*,M_2^*\rangle=\frac{1}{Z(L,M_1^*,M_2^*)}\sum_{P}{\sum}'_{N_1,N_2,N_4,N_5}
		|\underbrace{v_1v_1...}_{N_1}|\underbrace{v_2v_2...}_{N_2}|\underbrace{v_4v_4...}_{N_4}|\underbrace{v_5v_5...}_{N_5}\rangle,
		\label{lm1m3so5}
	\end{equation}
	where the sum $\sum_{P}$ is taken over all permutations $P$ for a given partition $\{N_1, N_2, N_4, N_5\}$, and the sum ${\sum}'_{N_1, N_2, N_4, N_5}$ is taken over all possible values of $N_1$, $N_2$, $N_4$ and $N_5$, subject to the constraints $N_2=M_1$, $2N_4+N_5=M_2^*$ and $N_1+N_2+N_4+N_5=L$.
	Here $N_1, N_2$, $N_4$ and $N_5$ denote the numbers of the lattice sites in the configurations $|v_1\rangle$, $|v_2\rangle$, $|v_4\rangle$ and $|v_5\rangle$, respectively, and $Z(L,M_1^*,M_2^*)$ is introduced to ensure that $|L,M_1^*,M_2^*\rangle$ is normalized. It takes the form
	\begin{equation}
		Z(L,M_1^*,M_2^*)=M_1^*!M_2^*!\frac{1}{\sqrt{2^{M_2^*}}}\sqrt{{\sum}'_{N_4}\frac{1}{4^{N_4}}C_{L}^{N_4}C_{L-N_4}^{M_2^*-2N_4}C_{L-M_2^*+N_4}^{M_1^*}}.
	\end{equation}

For the ${\rm SO}(6)$  ferromagnetic model, there are seven choices for the three commuting lowering operators: $F_1$, $F_2$ and $F_3$, $F_1$, $F_2$ and $F_4$, $F_1$, $F_2$ and $F_6$, $F_1$, $F_3$ and $F_4$, $F_1$, $F_4$ and $F_5$,  $F_1$, $F_5$ and $F_6$, and $F_2$, $F_3$ and $F_4$.
Here, we choose $F_1$, $F_2$ and $F_3$ to generate the degenerate ground states $|L,M_1,M_2,M_3\rangle$ ($M_1=0$, \ldots, $L$, $M_2=0$, \ldots, $L$, $M_3=0$, \ldots, $L$).
The fillings are defined as $f_1=M_1/L, f_2=M_2/L$, and $f_3=M_3/L$.

The four local orthonormal states $|v_2\rangle=-\sqrt{2}/2(0,  \mathrm{i},  0, 0, 1, 0)^T$, 
$|v_3\rangle=-\sqrt{2}/2(\mathrm{i},  0,  0, 0, 0, 1)^T$,  
$|v_4\rangle=\sqrt{2}/2(0, -\mathrm{i},  0, 0, 1, 0)^T$ and 
$|v_5\rangle=\sqrt{2}/2(0, 0, 1, -\mathrm{i}, 0, 0)^T$ are generated by applying the lowering operators $F_{1,j}$, $F_{2,j}$ and $F_{3,j}$ on  $|v_1\rangle_j$, i.e., 
$F_{1,j}|v_1\rangle=|v_2\rangle$,
$F_{2,j}|v_1\rangle=|v_3\rangle$,
$F_{3,j}|v_1\rangle=|v_4\rangle$,
$F_{1,j}F_{3,j}|v_1\rangle=|v_5\rangle$. 
Meanwhile, they are the eigenvectors of the local components of the Cartan generators: $H_{1,j}|v_2\rangle=-1/2|v_2\rangle$, $H_{2,j}|v_2\rangle=0|v_2\rangle$, $H_{3,j}|v_2\rangle=1/2|v_2\rangle$,   $H_{1,j}|v_3\rangle=0|v_3\rangle$, $H_{2,j}|v_3\rangle=-1/2|v_3\rangle$, $H_{3,j}|v_3\rangle=0|v_3\rangle$,   $H_{1,j}|v_4\rangle=1/2|v_4\rangle$, $H_{2,j}|v_4\rangle=0|v_4\rangle$, $H_{3,j}|v_4\rangle=-1/2|v_4\rangle$,  and  $H_{1,j}|v_5\rangle=-1/2|v_5\rangle$, $H_{2,j}|v_5\rangle=-1/2|v_5\rangle$, $H_{3,j}|v_5\rangle=-1/2|v_5\rangle$. 
In this way a sequence of the degenerate ground states $|L,M_1,M_2,M_3\rangle$ are generated from the repeated action of the lowering operators $F_1$, $F_2$ and $F_3$ on the highest weight state  $|{\rm hws}\rangle=|\otimes_{k=1}^{L} \{v_1\}_k\rangle$. Thus   $|L,M_1,M_2,M_3\rangle=1/{Z(L,M_1,M_2,M_3)}F_1^{M_1}F_2^{M_2}F_3^{M_3}|{\rm hws}\rangle$ may be rewritten as
	\begin{equation}
		|L,M_1,M_2,M_3\rangle=\frac{1}{Z(L,M_1,M_2,M_3)}\sum_{P}{\sum}'_{N_1,N_2, N_3,N_4,N_5}
		|\underbrace{v_1v_1...}_{N_1}|\underbrace{v_2v_2...}_{N_2}|\underbrace{v_3v_3...}_{N_3}
		|\underbrace{v_4v_4...}_{N_4}|\underbrace{v_5v_5...}_{N_5}\rangle \,.\label{ls1s2s3s4s5}
	\end{equation}
The sum $\sum_{P}$ is taken over all permutations $P$ for a given partition $\{N_1,N_2, N_3, N_4,N_5\}$, and the sum ${\sum}'_{N_1,N_2, N_3,N_4,N_5}$ is taken over all possible values of $N_1,N_2, N_3$, $N_4$ and $N_5$, subject to the constraints $N_2+N_5=M_1$, $N_4=M_2$, $N_3+N_5=M_3$, and $N_1+N_2+N_3+N_4+N_5=L$.
Here $N_1$, $N_2$, $N_3$, $N_4$ and $N_5$ denote the respective numbers of the lattice sites in the configurations $|v_1\rangle$, $|v_2\rangle$, $|v_3\rangle$,  $|v_4\rangle$ and $|v_5\rangle$.  $Z(L,M_1,M_2,M_3)$ is introduced to ensure that $|L,M_1,M_2,M_3\rangle$ is normalized, taking the form
\begin{equation}
	Z(L,M_1,M_2,M_3)=M_1!M_2!M_3!\sqrt{{\sum}'_{N_5}C_{L}^{N_5}C_{L-N_5}^{M_1-N_5}C_{L-M_1}^{M_2-N_5}C_{L-M_1-M_2+N_5}^{M_2}} \,.
\end{equation}

\section{Schmidt decomposition}\label{svdsm}

For the SO($2s+1$) ferromagnetic model under investigation, the exact Schmidt decomposition of the 
orthonormal basis states $| \, L,M_1,\ldots,M_r\rangle$ discussed in the previous section is of the form
\begin{equation}
		| \, L,M_1,\ldots,M_r\rangle= \sum_{k_1=0}^{\min(M_1,n)}\cdots\sum_{k_r=0}^{\min(M_r,n)}\lambda(L,n,k_1,\ldots,k_r,M_1,\ldots,M_r)
		\, | \, n,k_1,\ldots,k_r\rangle \, | \, L-n,M_1-k_1,\ldots,M_r-k_r\rangle \,. \label{svd}
\end{equation}
The Schmidt coefficients $\lambda(L,n,k_1,\ldots,k_r,M_1,\ldots,M_r)$ are given by
	\begin{equation}
		\lambda(L,n,k_1,\ldots,k_r,M_1,\ldots,M_r)
		=\prod_{\alpha=1}^rC_{M_\alpha}^{k_\alpha}\frac{Z(n,k_1,\ldots,k_r)Z_2(L-n,M_1-k_1,\ldots,M_r-k_r)}{Z_2(L,M_1,\ldots,M_r)} \,.
		\label{Lamm1mr}
	\end{equation}
		
For the orthonormal basis states $| \, L,M_1,\ldots,M_r\rangle$, the entanglement entropy $S_{\!L}(n,M_1,\ldots,M_r)$ for a block, consisting of $n$ contiguous lattice sites, follows from
\begin{align}
	S_{\!L}(n,M_1,\ldots,M_r)=-\sum_{k_1=0}^{\min(M_1,n)}\cdots\sum_{k_r=0}^{\min(M_r,n)}\Lambda(L,n,k_1,\ldots,k_r,M_1,\ldots,M_r)
		\log_{2}\Lambda(L,n,k_1,\ldots,k_r,M_1,\ldots,M_r),
		\label{snkm1m2m3}
\end{align}
where $\Lambda(L,n,k_1,\ldots,k_r,M_1,\ldots,M_r)$ are the eigenvalues of the reduced density matrix $\rho_L(n,M_1,\ldots,M_r)$, given by $\Lambda(L,n,k_1,\ldots,k_r,M_1,\ldots,M_r)=[\lambda(L,n,k_1,\ldots,k_r,M_1,\ldots,M_r)]^2$. 

Specializing to the SO(5) and SO(6) cases, the exact Schmidt decomposition is performed for the orthonormal basis states $|\, L,M_1,M_2\rangle$, $|\, L,M_1^*,M_2^*\rangle$ and  $| \, L,M_1,M_2,M_3\rangle$. 
Accordingly, the entanglement entropy for a block consisting of $n$ contiguous lattice sites follows from the above formula.
	
\end{widetext}	

\section{Entanglement entropy, number of type-B Goldstone modes and fractal dimension}\label{NBentropy}
 
For the orthonormal basis states (\ref{lmnr}) arising from SSB with type-B GMs, 
the entanglement entropy $S_{\!\!f}(L,n)$ is of the form~\cite{FMGM,finitesize}
\begin{equation}
	S_{\!\!f}(L,n)=\frac{N_B}{2} \log_2\frac{n(L-n)}{L} +S_{\!\!f0}.
	\label{slnf}
\end{equation}
The subscript $f$ refers to a set of fillings $f_1$, \ldots, $f_r$, defined as $f_1=M_1/L$, \ldots, $f_r=M_r/L$, and $S_{\!\!f0}$ is an additive non-universal constant.

Before proceeding, we emphasize that the finite system-size scaling~(\ref{slnf}) of the entanglement entropy $S_{\!\!f}(L,n)$ becomes a logarithmic scaling function $S_{\!\!f}(n)$ with the block size $n$ in the thermodynamic limit $L \rightarrow \infty$. This is consistent with a generic but heuristic argument~\cite{FMGM,finitesize}. That is, we have
\begin{equation}
S_{\!\!f}(n)=\frac{N_B}{2}\log_2n+S_{\!\!f0}.
\label{scaling}
\end{equation}
In addition, a field-theoretic prediction has been made by Castro-Alvaredo and Doyon~\cite{doyon} for the entanglement entropy of a linear combination on any support 
\begin{equation}
S_{\!\!f}(n)=\frac{d_f}{2}\log_2n+S_{\!\!f0}, \label{sr2}
\end{equation}
where $d_f$ is the fractal dimension of the support. This scaling relation (\ref{sr2}) was partially confirmed  for the SU(2) spin-$1/2$ ferromagnetic Heisenberg model in Ref.~\cite{doyon}. An alternative approach developed in Ref.~\cite{cantor}  
leads to the conclusion that this is the case for any quantum many-body systems undergoing SSB with type-B GMs. A comparison between the two scaling relations (\ref{scaling}) and (\ref{sr2}) shows that $d_f$ is identical to the number of type-B GMs, namely $d_f=N_B$, as far as the orthonormal basis states are concerned. 

We now make use of the results obtained in Sections \ref{statesvd} and \ref{svdsm} to test the validity
of the scaling forms (\ref{slnf}) and (\ref{scaling}) for the ferromagnetic SO(5) and SO(6) models as illustrative examples.

\section{Results for SO(5) and SO(6)}\label{results}

For the ferromagnetic SO(5) and SO(6) models, the subscript $f$ appearing in the entanglement entropy specifically refers to $f_1$ and $f_2$ or $f_1^*$ and $f_2^*$ for the ${\rm SO}(5)$ model and $f_1, f_2$, $f_3$ for the ${\rm SO}(6)$ model.

\begin{figure}[ht!]
	\centering
	\includegraphics[width=0.38\textwidth]{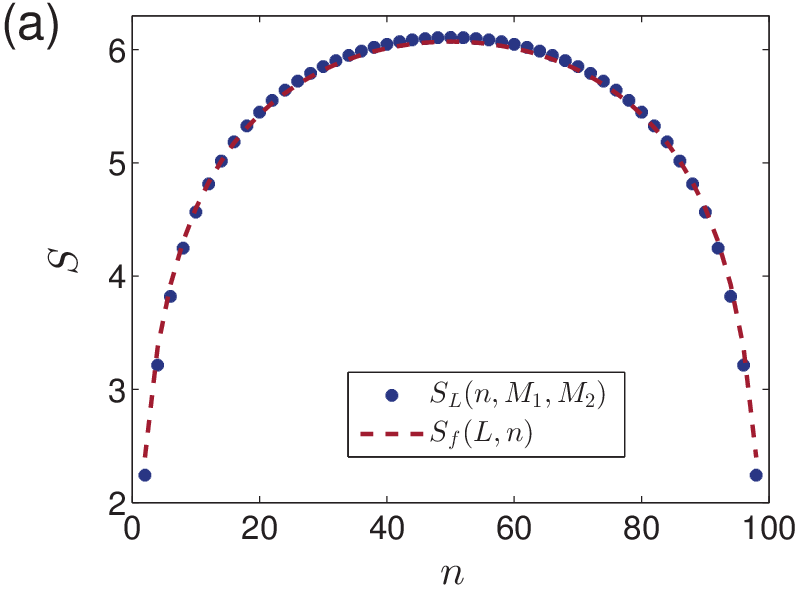}
	\includegraphics[width=0.38\textwidth]{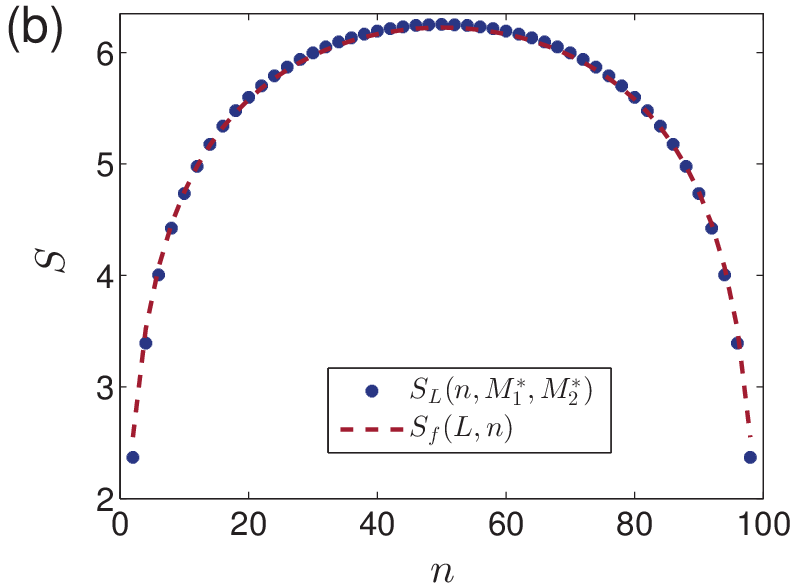}
\caption{ The ${\rm SO}(5)$ ferromagnetic model entanglement entropy (a) $S_{\!L}(n,M_1,M_2)$ versus $n$  for $|L,M_1,M_2\rangle$ with  $f_1=1/5$ and $f_2=1/5$ when $L=100$, (b) $S_{\!L}(n,M_1^*,M_2^*)$ versus $n$ for  $|L,M_1^*,M_2^*\rangle$ with  $f_1^*=1/4$ and $f_2^*=1/4$ when $L=100$. Shown for comparison in each case is the universal finite-size scaling $S_{\!\!f}(L,n)$ versus $n$. 
The best fits give (a) $S_{\!\!f0}=1.428$ and (b) $S_{\!\!f0}=1.580$.}
\label{so5com}
\end{figure}

For the ${\rm SO}(5)$ ferromagnetic model, Fig.~\ref{so5com} shows plots of the entanglement entropy 
versus $n$ for system size $L=100$ against the universal finite-size scaling function $S_{\!\!f}(L,n)$ 
versus block size $n$ with $n$ ranging from 10 to 90.
Our numerical data for the entanglement entropy is seen to fall on the curve $S_{\!\!f}(L,n)$, with the relative errors being less than $1.5\%$. 
Here and hereafter, we have regarded $S_{\!\!f}(L,n)$ as a function of $n$ for fixed $L$ and $f$.

\begin{figure}[ht!]
	\centering
	\includegraphics[width=0.38\textwidth]{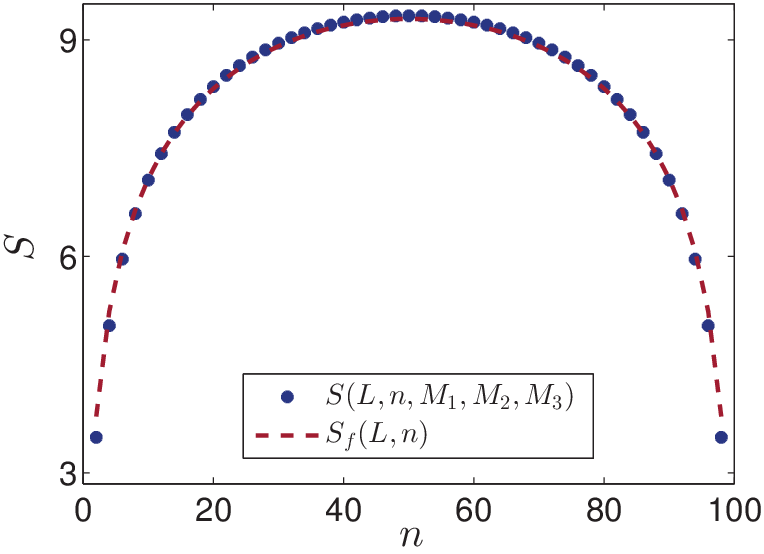}
	\caption{The ${\rm SO}(6)$ ferromagnetic model entanglement entropy $S_{\!L}(n,M_1,M_2,M_3)$ versus $n$ for $|L,M_1,M_2,M_3\rangle$ with the fillings $f_1=1/4$,  $f_2=1/4$ and $f_3=1/4$, when $L=100$ against the universal finite-size scaling $S_{\!\!f}(L,n)$ versus $n$. The best fitting gives  $S_{\!\!f0}=2.326 $.}
	\label{so6com}
\end{figure}

For the ${\rm SO}(6)$ ferromagnetic model, Fig.~\ref{so6com} shows plots of the entanglement entropy versus $n$ for system size $L=100$ against the universal finite-size scaling $S_{\!\!f}(L,n)$ versus $n$ with $n$ ranging from 10 to 90. 
The numerical data for the entanglement entropy is again seen to fall precisely on the curve $S_{\!\!f}(L,n)$.

\begin{figure}[ht!]
	\centering
	\includegraphics[width=0.38\textwidth]{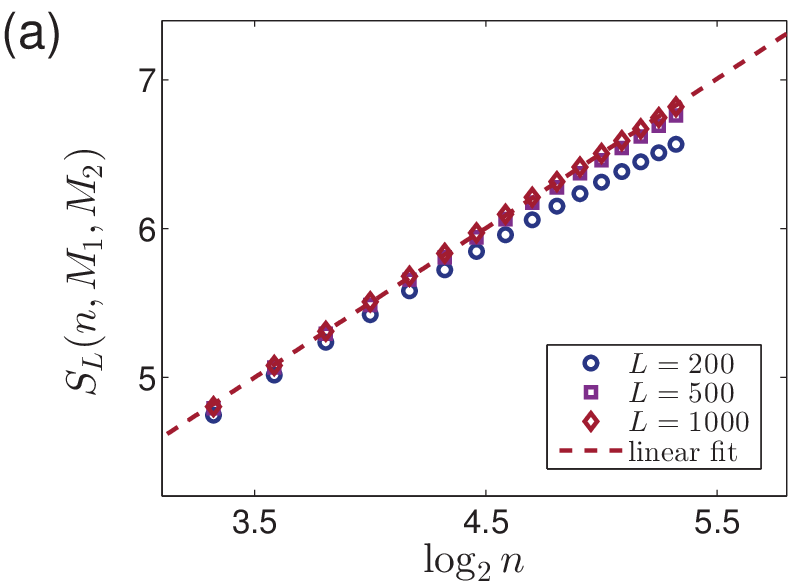}
	\includegraphics[width=0.38\textwidth]{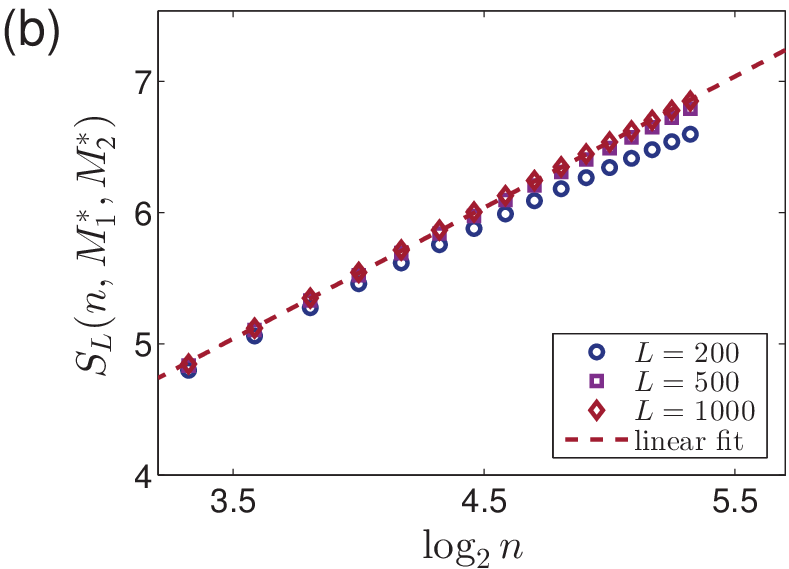}
	\caption{ The ${\rm SO}(5)$ ferromagnetic model entanglement entropy with increasing $L$ values:  (a) $S_{\!L}(n,M_1,M_2)$ versus $\log_2 n$, with  $M_1=L/5$ and $M_2=L/4$. (b) $S_{\!L}(n,M_1^*,M_2^*)$ versus $\log_2 n$, with $M_1^*=L/4$ and $M_2^*=L/4$.
	}
	\label{So5three}
\end{figure}

\begin{figure}[t]
	\centering
        \vskip 5mm
	\includegraphics[width=0.38\textwidth]{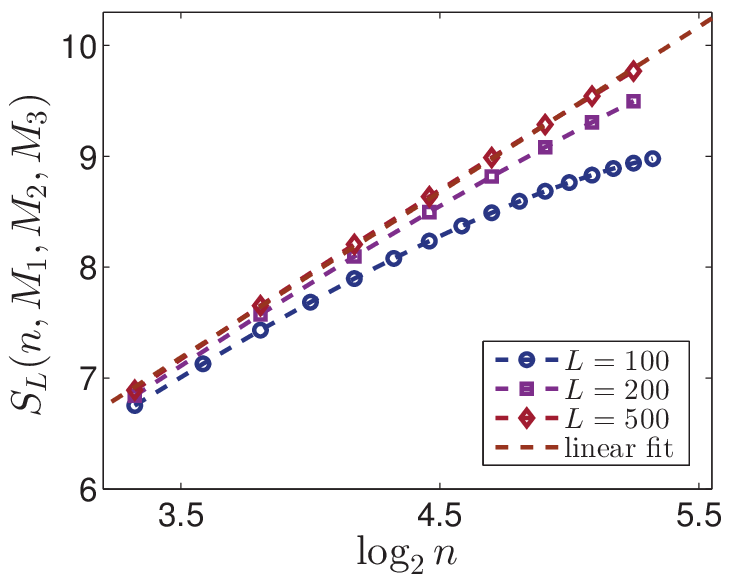}
	\caption{ The ${\rm SO}(6)$ ferromagnetic model entanglement entropy $S_{\!L}(n,M_1,M_2,M_3)$ versus $\log_2 n$ with increasing $L$ values. Here $M_1=L/4$, $M_2=L/4$ and $M_3=L/4$.
    The prefactor is close to $3/2$, within an error less than 1$\%$ when $L=500$.	}
	\label{So6two}
\end{figure}

In order to confirm how the logarithmic scaling behaviour of the entanglement entropy emerges in the thermodynamic limit, we show further plots in Fig.~\ref{So5three} and Fig.~\ref{So6two} with increasing $L$.
In both plots, a significant deviation from the logarithmic scaling behaviour is observed when $L$ is relatively small, but tends to vanish as $L$ increases.
Indeed, for both models, the prefactor is close to the exact value $1$, with an error less than 1.5$\%$ when $L=1000$.
For Fig.~\ref{So5three} we have (a) $S_{\!L}(n,M_1,M_2)=0.999\log_2n+1.480$ and (b) $S_{\!L}(n,M_1^*,M_2^*)=0.999\log_2n+1.542$.
For Fig.~\ref{So6two}, $S_{\!L}(n,M_1,M_2,M_3)=1.490\log_2n+1.970$.

\section{Conclusion}\label{summary}

We have systematically investigated SSB with type-B GMs in the SO($2s+1$) ferromagnetic model, with a focus on the ${\rm SO}(5)$ and  ${\rm SO}(6)$ models as illustrative examples. For this purpose, we have performed a finite system-size scaling analysis of the entanglement entropy for the orthonormal basis states in the ground state subspace. As a result, the universal finite system-size scaling function~\cite{FMGM,finitesize} for quantum many-body systems undergoing SSB with type-B GMs has been tested. The highly degenerate ground states for the ${\rm SO}(5)$ and the ${\rm SO}(6)$ ferromagnetic models are seen to be scale invariant, but not conformally invariant.

As a common feature with other quantum many-body systems undergoing SSB with GMs, the orthonormal basis states are generated from the repeated action of the commuting lowering operators, with the number of operators on the highest weight state equal to the rank of the symmetry group. It is found that the degenerate ground states admit an exact Schmidt decomposition, exposing the self-similarities in real space for the orthonormal basis states and revealing  an abstract fractal underlying the ground state subspace. As a consequence, the entanglement entropy scales logarithmically with the block size $n$ in the thermodynamic limit, with half the number of type-B GMs as 
the logarithmic prefactor for the orthonormal basis states. 
Combining with the field-theoretic result~\cite{doyon} and an alternative approach developed in \cite{cantor} allows identification of the number of type B GMs with fractal dimension, thus confirming the recent theoretical prediction for the orthonormal basis states~\cite{FMGM}. For the SO($2s+1$) ferromagnetic model under investigation here, $N_B = d_f = r$, with rank $r=s$ for integer $s$ and $r=s+1/2$ for half-odd-integer $s$. 
This result was explicitly confirmed for the spin $s=2$ and $s=5/2$ models for which we identified 
$N_B = d_f = 2$ for the SO(5) model and $N_B = d_f = 3$ for the SO(6) model.

The confirmation in the present study of the universal scaling form (\ref{scaling}) for the entanglement entropy adds to the previous demonstrations~\cite{FMGM,LLspin1,golden,SU4,finitesize} for the ferromagnetic SU(2) spin-$s$ Heisenberg, SU($N$+1) , staggered SU(3) spin-1 biquadratic, and staggered SU(4) spin-orbital models. Universal finite-size scaling and entanglement entropy for a type of scale-invariant states in two spatial dimensions and beyond have also been investigated~\cite{2d}. 

The results obtained here for the SO($2s+1$) ferromagnetic model also provide further information for the phase diagrams of the more general SO($2s+1$) bilinear-biquadratic spin chains. 
Specifically, the ferromagnetic ground state degeneracies obtained here at the value $\theta = \pi$ for the SO(5) and SO(6) cases are expected to persist throughout the ferromagnetic regime 
${\pi}/{2} < \theta < \pi+ \tan^{-1}[1/(2s-1)]$. 
These ground state degeneracies change however, at the two endpoints: $\theta= {\pi}/{2}$, corresponding to the staggered SU($2s+1$) ferromagnetic biquadratic models, and $\theta = \pi+ \tan^{-1}[1/(2s-1)]$, corresponding to the uniform SU($2s+1$) ferromagnetic bilinear-biquadratic models.
Moreover, at $\theta= {\pi}/{2}$ the ground state degeneracies are boundary condition dependent and exponentially increasing with the system size~\cite{golden,SU4}.
The observed SSB with type-B GMs also persists throughout the ferromagnetic regime, with the number of type-B GMs also changing at $\theta= {\pi}/{2}$ and $\theta = \pi+ \tan^{-1}[1/(2s-1)]$. 
For both endpoints $N_B = 2s$~\cite{FMGM,golden,SU4}. 

Most recently it has been shown that the exponential ferromagnetic ground state degeneracies imply the emergence of Goldstone flat bands, and a connection with quantum many-body scars is revealed~\cite{flatband}. 
Such developments may also be explored in the family of SO($2s+1$) ferromagnetic models.

\section{Acknowledgements}

We thank John Fjaerestad and Jon Links for their helpful comments and suggestions during the preparation of the manuscript.
I. P. M. acknowledges funding from the National Science and Technology Council (NSTC) Grant No. 112-2811-M-007-044 and 113-2112-M-007-MY2.

%%%%%%%%%%%%%%%%%%%%%%%%%%%%%%%%%%%%%%%%%%%%%%%%%Appendix%%%%%%%%%%%%%%%%%%%%%%%5
\onecolumngrid
\section*{Appendix}
\twocolumngrid
\setcounter{page}{1}

\subsection{Connection between the two choices for the generators of the symmetry groups ${\rm SO}(5)$  and ${\rm SO}(6)$}

For completeness, we outline the connection between the two choices for the generators of the symmetry groups ${\rm SO}(5)$  and ${\rm SO}(6)$: one set of the generators are $L_{ab}$ and another set of the generators are $H_\alpha$,  $E_\alpha$ and  $F_\alpha$, in addition to $E_\beta$ and  $F_\beta$.

For the symmetry group ${\rm SO}(5)$, the two choices of the generators of the symmetry groups ${\rm SO}(5)$ are connected as follows: 
$	H_1=(L_{15}+L_{24})/2$, $H_2=(-L_{15}+L_{24})/2$,	
$   E_1=(-L_{12}-L_{45}-\mathrm{i}L_{14}+\mathrm{i} L_{25})/2$,
$	E_2=(-L_{12}+L_{45}-\mathrm{i}L_{14}-\mathrm{i}L_{25})/2$, 
$   E_3=(L_{23}-\mathrm{i}L_{34})/2$,
$	E_4=(L_{13}-\mathrm{i}L_{35})/2$,	
$   F_1=(-L_{12}-L_{45}+\mathrm{i}L_{14}-\mathrm{i} L_{25})/2$,
$	F_2=(-L_{12}+L_{45}+\mathrm{i}L_{14}+\mathrm{i} L_{25})/2$,
$	F_3=(L_{23}-\mathrm{i}L_{34})/2$	and $F_4=(L_{13}-\mathrm{i}L_{35})/2$.  
Here $\langle n_d|L_{ab}|n_c\rangle=-\mathrm{i}(\delta_{ac}\delta_{bd}-\delta_{ad}\delta_{bc})$, with $|n_j\rangle$ ($j=1$, \ldots, 5) the $j$-th eigenvectors of the spin-$2$ operator $S_j^z$.   
They satisfy the following commutation relations:
$[H_1,E_1]=E_1$, $[H_1,F_1]=-F_1$, $[E_1,F_1]=2H_1$, $[H_2,E_2]=E_2$, $[H_2,F_2]=-F_2$, $[E_2,F_2]=2H_2$,	$[H_2,E_1]=0$, $[H_2,F_1]=0$, $[H_1,E_2]=0$, $[H_1,F_2]=0$,	$[H_1,E_3]=E_3/2$, $[H_1,E_4]=E_4/2$, $[H_1,F_3]=-F_3/2$,
$[H_1,F_4]=-F_4/2$, $[H_2,E_3]=E_3/2$, $[H_2,F_3]=-F_3/2$,	
$[H_2,E_4]=-E_4/2$, $[H_2,F_4]=F_4/2$,
$[E_3,F_3]=1/2(H_1+H_2)$, $[E_4,F_4]=1/2(H_1-H_2)$,
$[E_1,E_2]=0$, $[F_1,F_2]=0$, $[E_1,F_2]=0$, $[F_2,E_1]=0$, 
$[E_1,E_3]=0$, $[E_1,E_4]=0$, $[E_2,E_3]=0$, $[H_1,F_2]=0$.

For the symmetry group ${\rm SO}(6)$, the two choices of the generators of the symmetry groups ${\rm SO}(5)$ are connected as follows
$H_1=(L_{25}+L_{34})/2$,	$H_2=(L_{16}+L_{34})/2$,
$H_3=(-L_{25}+L_{34})/2$, $E_1=(L_{23}+L_{45}+\mathrm{i} L_{24}- \mathrm{i} L_{35})/2$,
$E_2=(L_{13}+L_{46}+\mathrm{i} L_{14}-\mathrm{i} L_{36})/2$,
$E_3=(L_{23}-L_{45}+\mathrm{i} L_{24}+\mathrm{i} L_{35})/2$,
$E_4=(L_{13}-L_{46}+\mathrm{i} L_{14}+\mathrm{i} L_{36})/2$,
$E_5=(L_{12}-L_{56}+\mathrm{i} L_{15}+\mathrm{i} L_{26})/2$,
$E_6=(L_{12}+L_{56}+\mathrm{i} L_{15}-\mathrm{i} L_{26})/2$,
$F_1=(L_{23}+L_{45}-\mathrm{i} L_{24}+\mathrm{i} L_{35})/2$,
$F_2=(L_{13}+L_{46}-\mathrm{i} L_{14}+\mathrm{i} L_{36})/2$,
$F_3=(L_{23}-L_{45}-\mathrm{i} L_{24}-\mathrm{i} L_{35})/2$,
$F_4=(L_{13}-L_{46}-\mathrm{i} L_{14}-\mathrm{i} L_{36})/2$,
$F_5=(L_{12}-L_{56}-\mathrm{i} L_{15}-\mathrm{i} L_{26})/2$,
$F_6=(L_{12}+L_{56}-\mathrm{i} L_{15}+\mathrm{i} L_{26})/2$. 
Here $\langle n_d|L_{ab}|n_c\rangle=-\mathrm{i}(\delta_{ac}\delta_{bd}-\delta_{ad}\delta_{bc})$, with $|n_j\rangle$ ($j=1$, \ldots, 6) the $j$-th eigenvectors of the spin-$5/2$ operator $S_j^z$.   
The commutation relations take the form
$[H_1,E_1]=E_1$, $[H_1,F_1]=-F_1$, $[E_1,F_1]=2H_1$, 
$[H_2,E_2]=E_2$, $[H_2,F_2]=-F_2$, $[E_2,F_2]=2H_2$,
$[H_3,E_3]=E_3$, $[H_3,F_3]=-F_3$, $[E_3,F_3]=2H_3$, 
$[E_1,E_2]=0$, $[E_2,E_3]=0$, $[E_3,E_1]=0$, 
$[F_1,F_2]=0$, $[F_2,F_3]=0$, $[F_3,F_1]=0$, in addition to other commutation relations:
$[E_4,F_4]=2(H_2-H_1-H_3)$, $[E_6,F_6]=2(H_1-H_2)$, $[E_5,F_5]=2 H_3$,
$[H_1,E_j]=E_j/2$,  ($j$ = 2, 4, 5, 6), $[H_1,E_3]=0$, 
$[H_1,F_j]=-F_j/2$,  ($j$ = 2, 4, 5, 6),  $[H_1,F_3]=0$,
$[H_2,E_j]=E_j/2$  ($j$ = 1, 3, 6), $[H_2,E_5]=-E_5/2$,  $[H_2,E_4]=0$,
$[H_2,F_j]=-F_j/2$  ($j$ = 1, 3, 6), $[H_2,F_5]=F_5/2$,  $[H_2,F_4]=0$,
$[H_3,E_j]=E_j/2$ ($j$ = 2, 4), $[H_3,E_j]=-E_j/2$  ($j$ = 5,6),  $[H_3,E_1]=0$,
$[H_3,F_j]=F_j/2$  ($j$ = 2,4), $[H_3,F_j]=-F_j/2$  ($j$ = 5,6), $[H_3,F_1]=0$,
$[E_1,E_j]=0$ ($j$ = 2,3,4,5,6),  $[E_2,E_j]=0$  ($j$ = 1,3,4,6), 
$[E_3,E_j]=0$  ($j$ = 1,2,4), $[E_4,E_j]=0$  ($j$ = 1,2,3,5),
$[E_5,E_j]=0$  ($j$ = 1,4,6), $[E_6,E_j]=0$  ($j$ = 1,2,5),	
$[F_1,F_j]=0$  ($j$ = 2,3,4,5,6), $[F_2,F_j]=0$  ($j$ = 1,3,4,6),
$[F_3,F_j]=0$  ($j$ = 1,2,4), $[F_4,F_j]=0$  ($j$ = 1,2,3,5),
$[F_5,F_j]=0$  ($j$ = 1,4,6) and $[F_6,F_j]=0$  ($j$ = 1,2,5).

\begin{widetext}

\setcounter{equation}{0}
\renewcommand{\theequation}{B\arabic{equation}}

\subsection{The realization of the ${\rm SO}(5)$ ferromagnetic model in terms of spin-2 operators}

The spin-2 formulation of the ${\rm SO}(5)$ model (\ref{ham}) was written explicitly in Ref.~\cite{alet}. 
The Hamiltonian of the ${\rm SO}(5)$ ferromagnetic model under consideration is given by
\begin{align}
\mathscr{H}=-\sum_ j \left(-1-\frac{5}{6} {\bf S}_j \cdot {\bf S}_{j+1}
+\frac{1}{9}({\bf S}_j \cdot {\bf S}_{j+1})^2+\frac{1}{18}({ \bf S}_j\cdot { \bf S}_{j+1})^3\right),\label{so5spin}
\end{align}
where $\textbf{S}_j=(S^x_j,S^y_j,S^z_j)$, with $S^x_j$, $S^y_j$, $S^z_j$ the spin-2 operators acting at site $j$.
More generally, the spin-$s$ formulation of the ${\rm SO}(n)$ model (\ref{ham}) for $n=2s+1$ odd can be written, up to a constant, in terms of spin projectors~\cite{son}, which we write as 
\begin{align}
\mathscr{H}=-\sum_ j \left( \cos \theta \,\, \mathcal{P}_j +n \left[ (n-2) \sin \theta - \cos \theta \right] P_j^{(0)} \right),
\end{align}
where $\mathcal{P}_j = P_j^{(0)}-P_j^{(1)}+ P_j^{(2)} - \ldots + P_j^{(n-1)}$,
with
\begin{equation}
P_j^{(i)} = \prod_{k=0\ne i}^{2s} \frac{X_j - x_k} {x_i-x_k}   , 
\end{equation}
where $X_j = {\bf S}_j \cdot {\bf S}_{j+1}$ and $x_k = \frac12 k(k+1) - s(s+1)$.
In this way the model can readily be written in terms of spin-$s$ operators for integer $s$. 

\end{widetext}

\end{document}